# High-resolution Doppler-free molecular spectroscopy with a continuous-wave optical parametric oscillator

E. V. Kovalchuk, D. Dekorsy, A. I. Lvovsky, C. Braxmaier, J. Mlynek, A. Peters and S. Schiller*

*Fachbereich Physik, Universität Konstanz, M696, D-78457 Konstanz, Germany*

We present a reliable, narrow linewidth (100 kHz) continuous-wave optical parametric oscillator (OPO) suitable for high-resolution spectroscopy applications. The OPO is based on a periodically-poled lithium-niobate crystal and features a specially designed intracavity etalon which permits its continuous tuning and stable operation at any desired wavelength in a wide operation range. We demonstrate Doppler-free spectroscopy on a rovibrational transition of methane at 3.39 µm.

Considerable technological progress in the fabrication of periodically poled nonlinear crystals, durable optical coatings, ultra-stable solid-state pump lasers and OPO design has recently led to a new generation of continuous-wave optical parametric oscillators (cw OPOs) with single frequency output covering a very wide range of the near-infrared spectrum (0.8 – 4.0 µm). These OPOs can now provide substantial amounts (10 – 250 mW) of coherent radiation with narrow linewidth (< 160 kHz), continuous tunability (several GHz) and excellent long-term stability.[1-5] This unique combination of features makes them attractive and promising tools for molecular spectroscopy, especially in the wavelength region around 3 µm, where fundamental vibrational modes of molecular bonds containing hydrogen occur. Accordingly, several Doppler-limited spectroscopic applications of single frequency cw OPOs were demonstrated since 1998.[3-7] However, to our knowledge no experiments employing cw OPOs for Doppler-free spectroscopy have been reported yet. In this Letter, we describe a new OPO design specially developed for high-resolution spectroscopy and present its application to observe Doppler-free resonances in methane at 3.39 µm.

Until recently, one of the remaining challenges in using cw OPOs for spectroscopic applications has been to prevent gaps in the spectral coverage and thereby to guarantee access to every atomic and molecular transition of interest. In case of the singly-resonant OPOs with resonated pump (PR-SROs) employed by our group, such gaps were generally caused by imperfections of optical coatings and irregularities in the nonlinear crystal. The resulting spurious modulation of the wide OPO gain profile often lead to unpredictable mode hops between widely spaced longitudinal cavity modes. In order to avoid these problems, it is desirable to provide additional mode selection. To this end, we have departed from our previous semi-monolithic cavity design[1] to implement an extended cavity setup with a specially designed intracavity etalon (ICE). Unlike the etalons commonly used in single-frequency lasers and also a variety of singly-resonant OPOs[4,5,8] this ICE is located inside a cavity with two resonated waves, of which it should only affect one (signal) and not the other (pump). It is therefore antireflection coated for the pump wave and the performance of this coating is crucial.

The OPO arrangement is shown schematically in Fig. 1. A monolithic diode-pumped Nd:YAG ring laser with a linewidth < 5 kHz and 1 W single-frequency output power at 1064 nm is used as a pump source. The frequency can be tuned by up to 40 GHz (9 GHz without mode hops) by changing the laser crystal temperature. An optical isolator prevents backreflections from the OPO cavity. The OPO is based on a periodically poled lithium niobate (PPLN) crystal of 19 mm length and 0.5 x 50 mm aperture. The crystal contains 33 gratings with periods ranging from 28.98 to 30.90 µm. At crystal temperatures between 145 and 170°C they provide quasi-phase-matching for signal waves in the wavelength range 1.48-1.93 µm, corresponding to 2.35-3.75 µm for the idler. The crystal is temperature stabilized within ± 2 mK and can be translated to select one of the gratings.

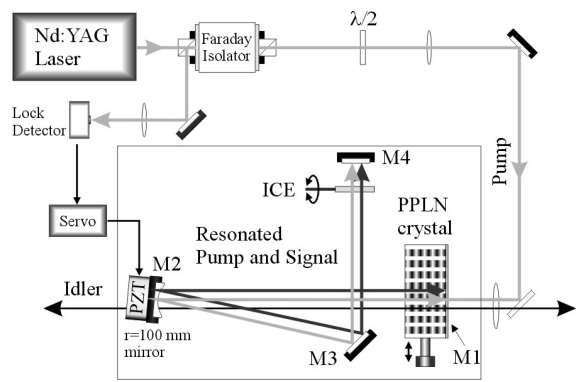

Fig.1. Schematic of the OPO setup (see text for details).

The extended OPO cavity consists of three plane and one concave mirrors. One mirror M1 is directly coated onto the plane entrance face of the nonlinear crystal and is highly reflecting for the pump (98%) and the signal (>99% for 1.35-1.65 µm, with a maximum of 99.9% at 1.5 µm), but highly transmitting (>98%) for the idler (3-4 µm). The concave mirror M2 of 100 mm curvature radius and the plane end mirror M4 have a reflectivity better than 99.8% for the pump and signal, and 1% for the idler. The plane mirror M3 minimizes the angle of incidence at the concave mirror and therefore astigmatic aberrations of the resonator mode. At 45° it reflects 99.8% of the pump and signal, and an average 3% of the idler. The second plane face of the crystal is antireflection coated for all three wavelengths with residual reflectivities of 0.15% for the pump, < 1% (1.35-1.65 μm, minimum 0.1% at 1.5 μm) for the signal and < 2% for the idler. Thus, the idler wave is not resonated and leaves the cavity through the mirrors M1 and M2.

The concave mirror is located 43 mm from the crystal surface and is mounted on a piezoelectric transducer (PZT). The optical cavity length is 420 mm, corresponding to a free spectral range (FSR) of 360 MHz. One waist of the fundamental cavity mode with $w_0 = 35$ μm (for the pump) is located at the entrance face of the crystal, another one with $w_0 = 440$ μm at the end mirror. The pump beam is focused into the PPLN crystal with an effective mode matching to the fundamental cavity mode of 99%. The pump and signal waves are both resonated in the same folded cavity, whose length is locked to the pump laser[1] using a Pound-Drever-Hall method.

To suppress mode hops caused by environmental perturbations and which were observed to span a range of 50 GHz an etalon is placed in the cavity. The ICE is made of a 1 mm thick undoped YAG plate (83 GHz FSR), coated to provide a reflectivity of 15% per surface for the signal. The coating exhibits a residual reflectivity of 0.75% for the pump, which still affects the threshold and tuning behavior of the OPO. However, it is low enough to warrant stable operation. To minimize walk-off losses, the ICE is placed near the cavity end mirror where the beam is almost collimated.

The minimum threshold pump power of the OPO was 305 mW at a signal wavelength of 1.55 μm. A finesse of 85 and a measured output transmission of 1.8% for the pump wavelength yield roundtrip cavity losses of 5.5%, of which 1% can be attributed to the etalon. At a pump input power of 808 mW the total idler output power (through M1 and M2) at 3.39 μm was 58 mW. The corresponding quantum efficiency of 23% is mainly limited by the nonoptimal transmission of the input mirror for the pump.[9]

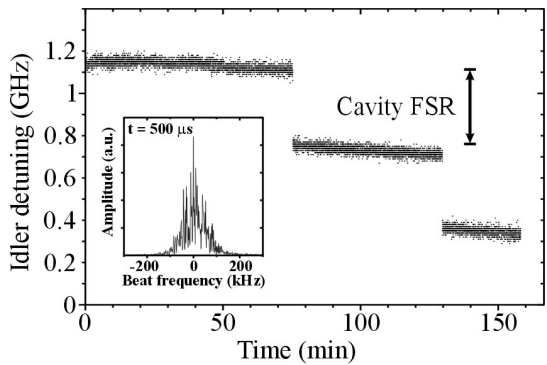

Fig.2. Long term frequency stability of the OPO output at 3.39 μm measured with a high resolution wavemeter (Burleigh WA-1500). The inset shows the measured OPO linewidth for an integration time of 500 μs.

We observed mode-hop free operation of the OPO over 75 min with a typical idler frequency drift of 0.5 MHz/min (Fig. 2). To determine the spectral characteristics of the OPO we performed a beat measurement between the idler wave and the narrow linewidth (50 Hz) radiation of a transportable He-Ne/CH₄ stabilized laser[10] at 3.39 μm. For an integration time of 500 μs we found a linewidth of ~100 kHz. For integration times longer than 10 ms the output linewidth becomes dominated by frequency jitter of the pump laser.

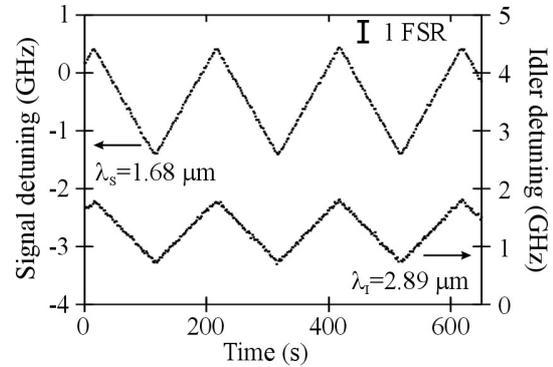

Fig.3. Continuous tuning of the OPO output over 1.8 GHz for the signal and 1 GHz for the idler, achieved by synchronously sweeping the pump frequency and tilting the ICE. For comparison, the FSR of the cavity is also displayed. Both frequencies were measured with a high resolution wavemeter.

By synchronously sweeping the pump frequency and tilting the ICE using a galvanometer scanner continuous tuning over 1.8 GHz for the signal (corresponding to 5 FSRs of the cavity) and 1 GHz for the idler waves was achieved (Fig. 3). The tuning range was limited by the PZT servo-amplifier and could be easily extended. In theory it should only be limited by the continuous tuning range of a pump laser.

1.

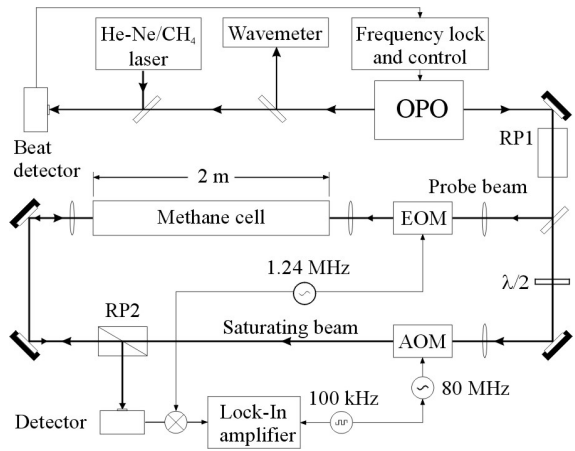

Fig.4. Doppler-free spectroscopy setup. One OPO idler output is split into the saturating and probe beams and directed through a methane cell. The probe beam is phase modulated by an electro-optic modulator (EOM) at a frequency of 1.24 MHz (modulation index ~1), the saturating beam is chopped by the acousto-optic modulator (AOM) at 100 kHz. After the cell the probe beam is sent to a fast photodetector, whose output is analyzed by a phase sensitive detection circuit and directed into a lock-in amplifier to extract the spectroscopic signal. The second OPO idler output is used for frequency monitoring by a wavemeter and frequency-locking to a He-Ne/CH₄ laser. RP1, RP2 - Rochon prism polarizers.

To demonstrate the capabilities of the new OPO, we have performed Doppler-free spectroscopy on the $F_2^{(2)}$ component of the P7 rovibrational transition of the methane molecule at 3.39 µm. This line is traditionally used as a high-stability reference line in He-Ne optical frequency standards. The spectroscopy setup is shown schematically in Fig. 4. The measurements were performed using the frequency modulation saturation technique[11] in a 2 m long methane cell. The diameter of the collimated counterpropagating probe (2 mW) and saturating (10 mW) beams inside the cell is about 8 mm, giving a transit-time broadening of approximately 30 kHz and a saturation parameter of ~ 0.1 at a methane pressure of 6 mTorr. Chopping of the saturating beam with an acousto-optic modulator (AOM) in combination with lock-in detection was used to eliminate the Doppler broadened background and spurious amplitude modulation. Using orthogonal polarizations for the pump and probe beams, together with the 80 MHz frequency shift provided by the AOM, helps very effectively to avoid perturbations of the OPO due to back reflections. To suppress slow frequency drifts of the OPO, the idler frequency was frequency-locked (bandwidth < 100 Hz) to the stabilized He-Ne laser with an offset of 40 MHz. It was then scanned linearly over the saturated resonance by changing the offset frequency.

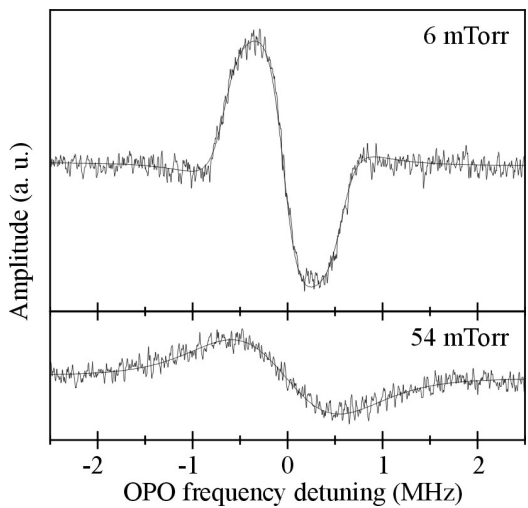

Fig.5. Doppler-free resonances of methane at 3.39 µm obtained by scanning the OPO idler frequency over 5 MHz in 50 s and using a lock-in time constant of 30 ms. Lines: fits according to a theoretical model.[11]

Doppler-free dispersion resonances observed at two different pressures are presented in Fig. 5. At 6 mTorr the linewidth of 500 kHz can be attributed to a combination of pressure broadening (200 kHz) and medium-term OPO frequency jitter (200-400 kHz). At 54 mTorr the observed linewidth of 1.1 MHz is dominated by pressure broadening.

In the future, we intend to further reduce the linewidth of the OPO by using a more rigid mechanical design and a higher bandwidth cavity lock. After these modifications we expect the linewidth of the OPO output radiation to replicate that of the pump laser (a few kHz). This would open up the route to ultra high-resolution spectroscopy of a multitude of much narrower Doppler-free spectral lines.

In particular, one could make use of transitions from low-lying rovibrational levels of the methane molecule which, at cryogenic temperatures, exhibit a spectroscopic signal which is by 2-3 orders of magnitude higher than that of the P7 line.[12] Using these lines would lead to frequency standards of significantly higher performance.

In conclusion, we have demonstrated a new design for a pump resonant SRO with a tunable intracavity etalon. The idler frequency is continuously tunable over 1 GHz by synchronously changing the pump laser frequency and tilting the etalon. An OPO output linewidth of 100 kHz was measured. The spectroscopic capability of the OPO has been demonstrated by recording Doppler-free resonances in methane.

We thank J.P. Meyn and R. Wallenstein for providing PPLN crystals. We are grateful to Mikhail Gubin and his team at the P.N. Lebedev Institute (Moscow, Russia) for making available the transportable He-Ne/CH$_4$ optical frequency standard. We acknowledge financial support for this work by the German Ministry of Education and Research, the German-Israeli Foundation, the Optik-Zentrum Konstanz, and the Gerhard-Hess Program of the German Science Foundation.

*Present address: Institut für Experimentalphysik, Universität Düsseldorf, D-40225 Düsseldorf, Germany